\documentclass[twocolumn,showpacs,preprintnumbers,amsmath,amssymb]{revtex4}

\usepackage{graphicx}
\usepackage{dcolumn}
\usepackage{bm}
\usepackage{epsfig}
\usepackage{color}

\begin{document}


\title{Eshelby description of highly viscous flow - half model, half theory}

\author{U. Buchenau}
 \email{buchenau-juelich@t-online.de}
\affiliation{%
Forschungszentrum J\"ulich GmbH, J\"ulich Centre for Neutron Science (JCNS-1) and Institute for Complex Systems (ICS-1),  52425 J\"ulich, GERMANY
}%

\date{July 4, 2018}

\begin{abstract}
A recent description of the highly viscous flow ascribes it to irreversible thermally activated Eshelby transitions, which transform a region of the undercooled liquid to a different structure with a different elastic misfit to the viscoelastic surroundings. The description is extended to include reversible Eshelby transitions, with the Kohlrausch exponent $\beta$ as a free parameter. The model answers several open questions in the field. 
\end{abstract}

\maketitle

\section{Introduction}

The undercooled liquids close to the glass transition are still an unsolved riddle \cite{bnap,angell,cavagna,bb,royall,lerner}.

The central question is the mechanism of the viscous flow. Its stretching shows that the process begins at relaxation times which are many decades shorter than the terminal relaxation time $\tau_c$. The density of these precursor processes in the barrier variable $v=\ln(\tau/\tau_c)$ increases with $\exp(\beta v)$ toward $v=0$. The Kohlrausch exponent $\beta$ is close to one half \cite{bnap,albena}.

It is obvious that the fast precursor processes must be back-and-forth jumps, while those close to $\tau_c$ must be irreversible no-return jumps. But there is no generally agreed theoretical picture of the crossover \cite{cavagna,bb,royall,lerner}, though one should mention Kia Ngai's coupling model \cite{ngai}, postulating a primitive relaxation five to six decades below $\tau_c$, which has stimulated much of the research in this field.

An equally important unsolved problem is the temperature dependence of the terminal relaxation time $\tau_c$, the so-called fragility, which seems to be linked with the Kauzmann paradoxon \cite{angell}. If one insists on describing it in terms of a thermally activated process, a jump over an energy barrier $V_c$, one has to postulate a $V_c$ which increases with decreasing temperature and even appears to diverge at the Kauzmann temperature.

Empirically, one finds in many glass formers $V_c\propto G$, where $G$ is the short time shear modulus \cite{nemilov,dyre,nelson,shov2015}. The explanation competes with the Adam-Gibbs explanation \cite{adam}, which ascribes the rise of $V_c$ to an increase in the size of the cooperatively rearranging regions, and which recently has also found experimental support in nonlinear dielectric data \cite{bauer}.

The present paper sheds some light on both problems, the stretching and the fragility. In addition, it provides useful recipes, not only for the fit of dynamical shear data, but also of dielectric, compressibility and particularly dynamical heat capacity data in undercooled liquids.

The paper is based on the recent description of the highly viscous flow in terms of irreversible Eshelby processes \cite{asyth}, which showed that the terminal relaxation time $\tau_c$ is a factor of eight longer than the Maxwell time $\tau_M=\eta/G$ ($\eta$ viscosity, $G$ short time shear modulus), and derived a well-defined relaxation time distribution for the irreversible structural relaxation processes.

But the exact results of the preceding paper \cite{asyth} are limited to the irreversible processes. The present paper extends this limited theoretical access to a full description of irreversible and reversible relaxation processes by a Kohlrausch Ansatz. The Ansatz connects irreversible processes at $v>0$ to reversible processes at $v<0$. Though still a model with one adaptable free parameter (the Kohlrausch $\beta$), it allows not only for satisfactory descriptions of experimental data, but also for considerable physical insight.

After this introduction, the paper presents the Kohlrausch Ansatz in Section II, adapting it to the earlier exact results \cite{asyth} for the irreversible processes. Section III applies the scheme to experimental shear, dielectric and heat capacity data in several glass formers. Section IV discusses the results and draws some conclusions.

\section{The model}

\subsection{The Kohlrausch Ansatz}

For a thermally activated structural rearrangement over an energy barrier $V$, the relaxation time $\tau$ is
\begin{equation}\label{arrh}
	\tau=\tau_0\exp(V/k_BT),
\end{equation}
with $\tau_0=10^{-13}$ s.

A structural rearrangement changes the shape of the region, changing the elastic shear misfit of the region to the viscoelastic surroundings. This implies that the transition is an Eshelby transition \cite{eshelby}.

In the following, let us describe the relaxation time dependence in terms of the barrier variable $v$
\begin{equation}\label{v}
	v=\frac{V-V_c}{k_BT}=\ln(\tau/\tau_c)
\end{equation}
which is zero at the terminal relaxation time $\tau_c$, with the corresponding energy barrier $V_c$.

In our Kohlrausch Ansatz, we describe both irreversible and reversible Eshelby transitions by a number density $n(v)$ of symmetric double well potentials (of course, the real transitions are asymmetric, so each symmetric double well potential stands for an integral over asymmetries)
\begin{equation}\label{kohl}
	n(v)=2(1-\beta)\exp(\beta v)
\end{equation}

The Kohlrausch Ansatz has not yet any theoretical justification \cite{cavagna,bb}, but is necessary to describe the slow rise on the right side of the peaks of frequency-dependent shear and dielectric \cite{bnap,albena} data.

The double-well potentials with $v>0$ are irreversible transitions, the rest reversible ones. For simplicity, we assume a sharp boundary between irreversible and reversible transitions here. A reversible transition makes back-and-forth jumps at its own short relaxation time $\tau=\tau_c\exp(v)$, until it decays with the longer terminal relaxation time $\tau_c$.

The jump rate into the other minimum of a symmetric double-well potential is $1/2\tau$. The lifetime of a region is found by the integration over all irreversible jumps
\begin{equation}
	\frac{1}{2\tau_c}\int_0^\infty\frac{1}{\exp(v)}n(v)dv=\frac{1}{\tau_c}.
\end{equation}

With the prefactor $2(1-\beta)$, the integral over all jumps for $\tau$ from $\tau_c$ to infinity provides indeed the decay rate $1/\tau_c$ for the regions.

For $\beta=1/2$, the total number of 2 of irreversible jumps equals the number of reversible transitions at shorter relaxation times. This will be used in the next subsection to determine the total recoverable compliance.

For the same argument, one needs the average rate $\overline{r}$ of the irreversible processes. At $v$, the decay rate $r=\exp(-v)/(2\tau_c)$ has to be weighted with its contribution $n(v)/\exp(v)$ to the total decay. Integrating the product over $v$, one gets the average rate 
\begin{equation}
	\overline{r}=\frac{1-\beta}{2\tau_c(2-\beta)},
\end{equation}
which is $1/6\tau_c$ for $\beta=1/2$.

This modifies eq. (6) of reference \cite{asyth}, the relation between the basic rate $r_0$ and the density $N_s$ of available structures in shear space, to
\begin{equation}
	r_0=\frac{N_s(1-\beta)}{2\tau_c(2-\beta)},
\end{equation}
because the average barrier for the jumps is higher than the one at $\tau_c$. 

Using eq. (8) of reference \cite{asyth}, this implies
\begin{equation}\label{ns}
	N_s=\frac{2-\beta}{16(1-\beta)}.
\end{equation}

This is important, because $N_s$ determines the density of reversible Eshelby states at the crossover. 

\subsection{Connecting viscous decay and reversible transitions}

As derived in the preceding paper \cite{asyth}, $\tau_c$ is only the average decay time; the precise decay time depends on the shear stress state of the region. Strongly stressed regions decay faster than a stress-free one.

A measure for the shear stress is the dimensionless quantity
\begin{equation}
	e^2=\frac{E_{shear}}{k_BT}.
\end{equation}
The shear stress energy $E_{shear}$ comes from the elastic shear misfit between the region and the viscoelastic surroundings, which causes an elastic deformation of both the region and the surroundings.  About half of it is shear energy of the region. The other half is shear energy of the surroundings  \cite{eshelby}.

As shown in the preceding paper \cite{asyth}, a strongly strained region with a high $e^2$ has much more possibilities to make an irreversible jump than a weakly strained one. Quantitatively 
\begin{equation}\label{plt}
	\frac{4\sqrt{2}\tau_c}{\tau}=\exp(e^2/2).
\end{equation}
The longest decay time is $\tau=4\sqrt(2)\tau_c$ for an unstrained region.

This leads to a normalized distribution of the logarithm of the decay time \cite{asyth}
\begin{equation}\label{pt}
	l_{irrev}(v)=\frac{1}{3\sqrt{2\pi}}\exp(v^2)\left(\ln(4\sqrt{2})-v\right)^{3/2}.
\end{equation}

A strong argument for the validity of these considerations is the fit of dynamical heat capacity $c_p(\omega)$-data of a vacuum pump oil \cite{asyth} in terms of this relaxation time distribution, with only $\Delta c_p$ as a free parameter. $\tau_c=8\tau_M$ was taken from dynamic shear data of the same sample at the same temperature in the same cryostat.

In order to make use of the connectivity of reversible and irreversible transitions at $\tau_c$ in the Kohlrausch Ansatz of eq. (\ref{kohl}), one has to relate the density of double-well transitions to their shear compliance contribution.

For the irreversible transitions, this has been done in the preceding paper \cite{asyth}. Within the time $\tau_c=8\tau_M=8\eta/G$, their integrated contribution to the shear compliance is $8/G$.

The contribution of a reversible Eshelby transition to the shear compliance is smaller than the one of an irreversible Eshelby transition, for two reasons. 

The first has been already discussed in reference \cite{asyth}: in a sheared Eshelby state the shear stress energy is equally distributed between the region and the surroundings \cite{eshelby}. An irreversible jump relaxes both the region and the surroundings, because it does not return to restore the original state of the surroundings. So it relaxes the volume $2NV_p$ ($N$ number of particles, $V_p$ particle volume).

This is different for a reversible jump, for which the surroundings stay in a state adapted to the thermal average of the two states. Therefore the reversible transitions are a factor of two weaker to start with.

The second reason lies also in the different nature of the two processes. 

To see this, one has to remember the derivation of the average squared shear angle $\overline{\epsilon^2}$ for irreversible processes from reference \cite{asyth}, and then calculate the same quantity for the reversible processes.

Assume a region with a given shear state $e_0^2$. Its transition to another possible structure of the region with a different shear misfit $e^2$ gets \cite{asyth} a transition rate factor $\exp((e_0-e)^2/2)$. It is this transition rate factor which limits the average squared shear angle of the irreversible processes.

The average squared shear angle starting from $e_0^2$ within the time $\tau_c$ is proportional to 
\begin{equation}
	\overline{(e_0-e)^2}_{irrev}=\int_0^\infty (e_0^2+e^2)e^4\exp((e_0^2-e^2)/2)de.
\end{equation}

To get the total average squared shear angle, one has to integrate over all $e_0$ values with their thermal weight
proportional to $e_0^4\exp(-e_0^2)$.

For the reversible processes, the limiting factor is not the transition rate, but rather the asymmetry factor $1/\cosh^2((e_0^2-e^2)/2)$ from the energy difference of the two minima.

Thus for a reversible process one has the expectation value
\begin{equation}
	\overline{(e_0-e)^2}_{rev}=\int_0^\infty(e_0^2+e^2)e^4/\cosh^2((e_0^2-e^2)/2)de,
\end{equation}
which again has to be integrated over $e_0$. Unlike the integral over the irreversible processes, this integral is not easily solvable and has to be evaluated numerically.

The ratio of the two double integrals over $e$ and $e_0$ yields a second weakening factor $f_{r0}=0.4409$ for the reversible processes.

The irreversible processes provide the viscous shear compliance $8/G$ within the terminal relaxation time $\tau_c=8\eta/G$. For $\beta$ exactly 1/2, there is an equal number of transitions within $\tau_c$ for reversible and irreversible processes. Therefore the total contribution of the reversible processes to the recoverable compliance must be $8f_{r0}/2G=4f_{r0}/G$.

In terms of a density $l_r(v)$ of recoverable compliance processes, with a contribution $l_r(v)dv/G$ to the recoverable compliance in the interval from $v$ to $v+dv$, this requires an integral of $l_r(v)$ from $-\infty$ to zero of $4f_{r0}$, i.e. for $\exp(v/2)$ a prefactor of $2f_{r0}$.

For the general case of a $\beta$ not exactly 1/2, one has the prefactor $2(1-\beta)$ from eq. (\ref{kohl}). To this, one has to add a prefactor $(2-\beta)/3(1-\beta)$, which comes from the $\beta$-variation of eq. (\ref{ns}). Together
\begin{equation}\label{lv}
	l_r(v)=\frac{f_{r0}(8-4\beta)}{3}\exp(\beta v)\exp(-\exp(v)),
\end{equation}
where the last factor $\exp(-\exp(v))$ ensures the upper cutoff at $v=0$.

In substances with secondary relaxation peaks, one can add a gaussian distribution of additional barriers to $l_r(v)$.

Having $G$, $\tau_c=8\tau_M$ and $l_r(v)$, one can calculate the shear compliance $J(\omega)=1/G(\omega)$ from
\begin{equation}\label{jom}
	GJ(\omega)=1+\int_{-\infty}^\infty \frac{l_r(v)}{1+i\omega\tau_c\exp(v)}dv-\frac{i}{\omega\tau_M}.
\end{equation}

Note that here the relaxation time distribution of the irreversible processes, eq. (\ref{pt}), does not enter; altogether, they fix the viscosity $\eta$, which in turn fixes $\tau_M$ and $\tau_c$. 

This is different for other physical quantities, like the dielectric susceptibility, where the relaxation time distribution of the irreversible processes enters explicitly, as explained in the next subsection. 

\subsection{Dielectric and other data}

The quantitative relation between the strength of the irreversible and the reversible processes helps to find the appropriate decay function $p_\epsilon(v)$ for other physical variables like the dielectric susceptibility or the dynamic compressibility.

If the coupling constant for the other physical variable is proportional to the one for the shear at all relaxation times
\begin{equation}\label{peps}
p_\epsilon(v)=f_{norm}(8l_{irrev}(v)+l_r(v)),
\end{equation}
where $f_{norm}$ is a normalization factor, the density of irreversible processes $l_{irrev}(v)$ is given by eq. (\ref{pt}) and the density $l_r(v)$ of reversible processes is given by eq. (\ref{lv}).

In the next Section III, the comparison to experiment, we will see that one often needs an earlier cutoff. This can be done by multiplying eq. (\ref{peps}) with the factor $\exp(-\exp(v+v_\epsilon))$ and adapting the normalization factor.

A special case is the dynamic heat capacity, in which one does not expect to see the reversible processes, because they do not change the average structural energy. 

In fact, as mentioned above, the dynamic heat capacity in the vacuum pump oil PPE has been found to be well described \cite{asyth} by eq. (\ref{pt}) for the irreversible processes alone. In the next section, this will be corroborated by more examples. 

\section{Comparison to experiment}

\subsection{Overview}

If one applies the model description of the preceding Section II to different glass formers, one learns many things.

In the simplest case, the dynamic shear relaxation is described by the four parameters $G$, $\tau_c$, $\beta$ and $f_r$, replacing the theoretical value $f_{r0}$ in eq. (\ref{lv}) by the fit parameter $f_r$.

The simplest form of the model works in many glass formers, with $f_r$ equal to the theoretical prefactor $f_{r0}=0.4409$ within experimental error (see Table I). Some of these simple examples have covalent bonding, some metallic bonding and some van der Waals bonding. These simple glass formers will be discussed in detail in Section III. B.

But the model description in its simplest form does not work at all in glycerol and other hydrogen-bonded substances.

\begin{figure}   
\hspace{-0cm} \vspace{0cm} \epsfig{file=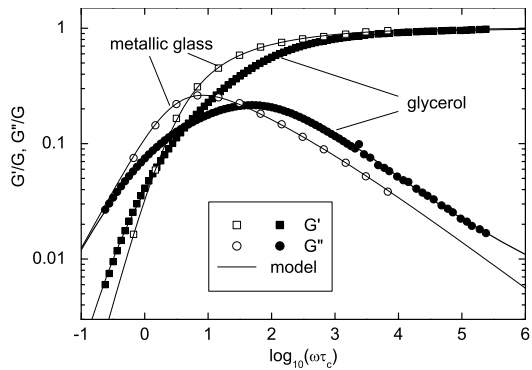,width=7 cm,angle=0} \vspace{0cm} \caption{The normalized shear relaxation curves of the metallic glass Zr$_{65}$Al$_{7.5}$Cu$_{17.5}$Ni$_{10}$ at 644.2 K \cite{schro} and of glycerol at 196 K \cite{glynew}. The continuous lines are fits in terms of the model with the parameters given in Table I and II.}
\end{figure}

This is illustrated in Fig. 1, which compares the normalized shear relaxation curves of the metallic glass Zr$_{65}$Al$_{7.5}$Cu$_{17.5}$Ni$_{10}$ \cite{schro} with those of the hydrogen bonding molecular glass former glycerol \cite{glynew}.

In these normalized curves, the shear modulus is the same, the viscosity is the same, and the terminal relaxation time is the same. Nevertheless, they look quite different, even on a logarithmic scale.

The metallic glass is very well described within the simplest form of the model, with $f_r=0.39$  and a Kohlrausch $\beta$ of 0.415 (see Table 1).

The same holds for the two other metallic glasses Pd$_{40}$Ni$_{40}$P$_{20}$ and vitralloy 4 in Table I. In fact, it is probably valid for all metallic glass formers, which have been shown to have very similar shear relaxation curves for a large number of examples \cite{met}.

In glycerol, the $f_r$ value is a factor of 3.4 larger than the theoretical value, one has to invoke a large return probability for the Eshelby transitions with relaxation times larger than $\tau_c$, and one has a hidden secondary relaxation peak (see Table II and the detailed discussion in Section III. D).

The model fails in the opposite direction in silica and some silica-rich glasses and minerals, where $f_r$ is about half of the theoretical value (see Table I). But there might be a physical reason for this failure, as will be discussed in Section III. B.

The most surprising result of the comparison to experiment is the effect of secondary relaxation peaks. Table II shows that $f_r$ increases with the increasing secondary peak amplitude, compatible with the interpretation that the faster relaxations begin to become a part of the slower ones and diminish their contribution to the viscous flow.

Section III. C discusses this secondary relaxation effect and its interpretation for two examples, dibutyl phtalate  and squalane. Section III. D deals with the hydrogen bonding glass formers.

\subsection{Simple glass formers}

\begin{table}[htbp]
	\centering
		\begin{tabular}{|c|c|c|c|c|c|c|}
\hline
subst.                                             &$T$    & $G$  &$\ln{\tau_c/s}$&  $\beta$   &  $f_r$   &$GJ_0$ \\
\hline   
                                                   &K      &GPa   &               &            &          &       \\ 
\hline
metallic glass formers                             &       &      &               &            &          &       \\
Zr$_{65}$Al$_{7.5}$Cu$_{17.5}$Ni$_{10}$            &644.2  &27.6  &4.232          &.415        &.39       &2.78   \\
Zr$_{65}$Al$_{7.5}$Cu$_{17.5}$Ni$_{10}$            &649.2  &28.6  &3.52           &.436        &.45       &2.94   \\
Pd$_{40}$Ni$_{40}$P$_{20}$                         &591.2  &31.7  &.92            &.415        &.45       &3.03   \\
Pd$_{40}$Ni$_{40}$P$_{20}$                         &596.2  &30.7  &.171           &.419        &.45       &2.97   \\
Pd$_{40}$Ni$_{40}$P$_{20}$                         &601.2  &32.7  &-.778          &.397        &.43       &3.08   \\
vitralloy-4                                        &618.2  &31.7  &6.725          &.431        &.52       &3.22   \\
\hline
van der Waals                                      &       &      &               &            &          &       \\
DC704                                              &214    &1.08  & 1.856         &.482        &.45       &2.68   \\
DC704                                              &224    &0.92  &-5.264         &.498        &.50       &2.80   \\
PPE                                                &250    &1.11  & 1.892         &.503        &.41       &2.43   \\
PPE                                                &264    &0.87  &-6.513         &.493        &.46       &2.67   \\
triphenylethylene                                  &268    &1.01  &-5.032         &.516        &.49       &2.68   \\
m-toluidine                                        &190    &1.10  &-0.712         &.433        &.47       &3.00   \\
\hline
networks                                           &       &      &               &            &          &       \\
B$_2$O$_3$                                         &650    &2.34  &-15.1          &1.03        &1.68      &3.16   \\ 
silica                                             &1449   &29.5  & 6.73          &.373        &.21       &2.08   \\
(SiO$_2$)$_{75}$(Na$_2$O)$_{25}$                   &752    &16.9  & 6.58          &.382        &.18       &1.91   \\
(SiO$_2$)$_{67}$(Na$_2$O)$_{33}$                   &728    &14.5  & 6.66          &.377        &.19       &1.95   \\
soda lime glass                                    &847    &25.8  & 2.68          &.449        &.43       &2.74   \\
rhyolite                                           &825    &31.1  & 1.90          &.299        &.26       &2.78   \\
haplogranite-F10                                   &986    &21.7  & 3.08          &.440        &.47       &2.98   \\
\hline
		\end{tabular}
	\caption{Model parameters fitted to shear relaxation data of glass formers without pronounced secondary relaxation peak: $G$ shear modulus, $\tau_c$ terminal relaxation time, $\beta$ Kohlrausch exponent and $f_r$ prefactor for the reversible relaxations. $GJ_0$ is the resulting recoverable compliance. References for the data: Zr$_{65}$Al$_{7.5}$Cu$_{17.5}$Ni$_{10}$ and Pd$_{40}$Ni$_{40}$P$_{20}$ \cite{schro}; vitralloy-4 \cite{pelletier}; DC704 \cite{tina}; PPE and triphenylethylene \cite{niss}; m-toluidine \cite{maggi}; B$_2$O$_3$ \cite{tauke}; silica, (SiO$_2$)$_{75}$(Na$_2$O)$_{25}$ and (SiO$_2$)$_{67}$(Na$_2$O)$_{33}$ \cite{mills}; soda lime glass \cite{donth2}; rhyolite \cite{webb1992}; haplogranite \cite{webb1993}.}
	\label{tab:rse1}
\end{table}

\begin{figure}   
\hspace{-0cm} \vspace{0cm} \epsfig{file=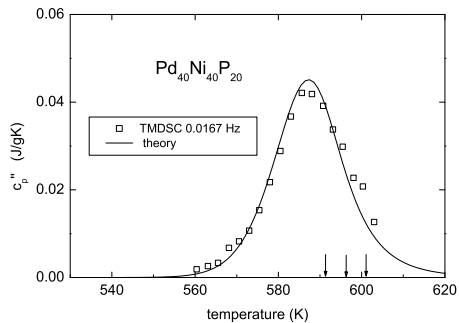,width=6 cm,angle=0} \vspace{0cm} \caption{TMDSC (Temperature Modulated Differential SCanning) data \cite{wilde} for the imaginary part $c_p''$ of the dynamic heat capacity of Pd$_{40}$Ni$_{40}$P$_{20}$. The continuous line is calculated from eq. (\ref{pt}) for the irreversible Eshelby processes, using $\Delta c_p=0.109$ J/gK and the $\tau_c$ temperature dependence of the three shear measurements in Table I. The three arrows show the temperatures of the shear data.}
\end{figure}

As shown in Table I, the theoretical value $f_{r0}=0.4409$ describes metallic glasses \cite{schro,pelletier} and van der Waals glass formers without strong secondary relaxation peak \cite{tina,niss,maggi} within experimental error.

The validity of the simple scheme for metallic glasses is independently demonstrated by the dynamic heat capacity data \cite{wilde} for Pd$_{40}$Ni$_{40}$P$_{20}$ shown in Fig. 2.

The temperature dependence of the dynamic heat capacity follows from the plausible assumption that it is exclusively due to the irreversible processes, using eq. (\ref{pt}) and taking $\tau_c$ as well as its Vogel-Fulcher temperature dependence from the fit of the shear data in Table I.

The finding corroborates the same finding in the vacuum pump oil PPE \cite{asyth}. Note that the model parameter $\beta$ is not involved; the explanation of the irreversible part of the decay in terms of Eshelby transitions \cite{asyth} is not a model, but an exact theory. Section III. D will show that the result holds also in hydrogen bonding substances.

The simple model fails in B$_2$O$_3$ \cite{tauke}, silica \cite{mills} and some silicates \cite{mills,donth2,webb1992,webb1993} (see Table I).

The reasons for the failure in B$_2$O$_3$ and silica are already well analyzed in the literature \cite{tauke,mott,kieffer}.

In the case of B$_2$O$_3$, the flow process rapidly loses its stretching with increasing temperature, transforming into a single Debye process above 800 K \cite{tauke}. This is consistent with the high Kohlrausch $\beta=1.03$ which one finds at 650 K.

With the loss of the stretching, the flow process also loses its fragility. Above 800 K, the temperature dependence follows an Arrhenius law \cite{tauke}, which has been interpreted as an exchange of two oxygens in neighboring covalent B-O bonds, disrupting and reforming boroxol rings \cite{kieffer}.

At 650 K, this process lies in the region of reversible processes. According to the theoretical considerations of the preceding Section II, it then loses most of its strength. The flow process should be taken over by a combination of two such oxygen exchange processes, a more cooperative process, but not yet fully describable in the scheme of the present paper.

The small value of $f_r$ in silica and some silicates requires a different explanation, but is most probably also due to covalent bond breaking. In fact, according to Mott's picture of the viscous flow in silica \cite{mott}, the flow is due to the creation and migration of broken Si-O bonds. In this picture, one would not expect any reversible Eshelby transitions, so $f_r=0$. 

The experimental result shows that this is not true; there are some reversible Eshelby transitions even in silica.

In fact, in this well-studied case one has a wealth of internal friction measurements in the glass phase \cite{philmag2002} and can pursue the reversible processes down to low temperatures. 

The internal friction measurements in the glass phase are conveniently expressed \cite{burel} in terms of the barrier density $f(V)$, where $f(V)dV=\delta G/G$ describes the reduction of the shear modulus by the relaxations with barriers between $V$ and $V+dV$.

Since the total reduction of $G$ by relaxations in the glass phase remains small, one can use $\Delta G/G=\delta JG$, where $\delta J$ is the increase of the shear compliance from the relaxations with barriers between $V$ and $V+dV$, together with the definition of $v$ in eq. (\ref{v}) and finds the Eshelby contribution
\begin{equation}
	f(V)=\frac{f_{r}(8-4\beta)}{3k_BT_g}\exp(\beta(V-V_c)/k_BT_g).
\end{equation}

The comparison in Fig. 3 shows that the reversible Eshelby relaxations exist also in the glass phase, down to about one fifth of the flow barrier of 4.573 eV, keeping to the same Kohlrausch $\beta$. If one would not see it, one would not believe it.

\begin{figure}   
\hspace{-0cm} \vspace{0cm} \epsfig{file=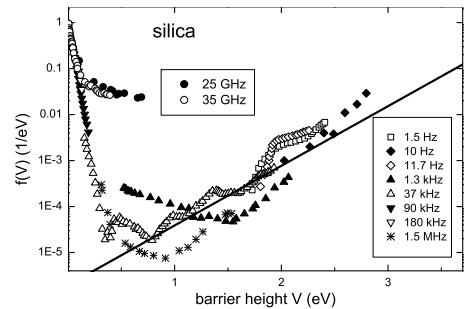,width=6 cm,angle=0} \vspace{0cm} \caption{Comparison of the fitted density of reversible Eshelby states to internal friction data in the glass phase of silica\cite{philmag2002}. The line is calculated from the parameters in Table I.}
\end{figure}

\begin{figure}   
\hspace{-0cm} \vspace{0cm} \epsfig{file=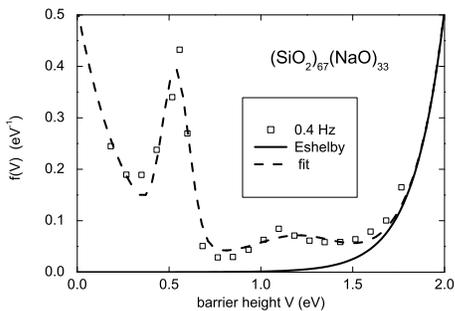,width=6 cm,angle=0} \vspace{0cm} \caption{Comparison of the fitted density of reversible Eshelby states to torsion pendulum data in the glass phase of (SiO$_2$)$_{75}$(Na$_2$O)$_{25}$ \cite{n25p4}. Continuous line calculated from the parameters in Table I.}
\end{figure}

The same comparison is done in Fig. 4  for torsion pendulum data \cite{n25p4} at 0.4 Hz in (SiO$_2$)$_{75}$(Na$_2$O)$_{25}$, with the extrapolation down from $V_c=2.236$ eV. 

Here, one sees the Eshelby transitions only at the upper end of the spectrum, because there are two pronounced secondary relaxation peaks. The rise of the spectrum at the upper end indicates a barrier density which is a factor 1.8 higher than the fit result of Table I, consistent with an $f_r$ of 0.33, closer to the theoretical value.

For (SiO$_2$)$_{67}$(Na$_2$O)$_{33}$, later measurements \cite{webb2010} showed that the original measurement in the melt \cite{mills} had a smaller modulus than the real one, indicating some gliding at the clamps holding the sample.

The torsion pendulum method shows the exponential rise toward high temperatures in many more network glasses \cite{rindone}.

The comparison to the internal friction in the glass supports the fitted $f_r$- and $\beta$-values for silica in Table I, but does not help to understand how $f_r$ can be a factor of two smaller than the theoretical value.

In order to explain that, one has to postulate bond breaking processes in the irreversible region, which lower the necessary amount of irreversible Eshelby transitions by about a factor of two. In fact, the estimated bond-breaking energy \cite{sucov} for an Si-O bond lies at 3.09 eV, even below the flow barrier.

In (SiO$_2$)$_{75}$(Na$_2$O)$_{25}$, (SiO$_2$)$_{67}$(Na$_2$O)$_{33}$ and in rhyolite, where the glass temperature lies about a factor of two lower than in silica, it is not sure whether one can use the same explanation or whether one should rather ascribe it to the difficulty of an accurate measurement of the shear relaxation. 

In the soda lime glass \cite{donth2} and in the haplogranite F-10 (a haplogranite with ten percent fluor \cite{webb1993}), one gets again the theoretical value of $f_r$.

\subsection{Secondary relaxation peaks}

\begin{table*}[htbp]
	\centering
		\begin{tabular}{|c|c|c|c|c|c|c|c|c|c|c|c|c|c|c|}
\hline
subst.&$T$&$G$&$\ln{\tau_c/s}$&$\beta$&$f_r$&$f_{r0}$&$p_r$&amp&pos&fwhm&$GJ_0$&$\Delta\epsilon$&$v_\epsilon$&$f_\epsilon$\\
\hline   
      &K    &GPa&        &       &     &        &   &     & eV  & eV   &      &                &            &      \\ 
\hline
van der Waals          &&&&&&&&&&&&&& \\
dibutyl phtalate &178  &0.90& 3.92 &.5   &.53     & .48&0.  &.0125&.27   &.108  &2.96& 7.2      &  0.13      &.93   \\
dibutyl phtalate &180  &0.87& 2.09 &.5   &.55     & .49&0.  &.0168&.27   &.108  &3.07&          &            &.     \\
dibutyl phtalate &182  &0.82& 0.39 &.5   &.56     & .49&0.  &.0202&.27   &.108  &3.15& 6.8      &  0.0       &.84   \\
dibutyl phtalate &184  &0.79&-1.17 &.5   &.58     & .49&0.  &.0234&.27   &.108  &3.21&          &            &      \\
squalane         &168  &1.27& 5.42 &.5   &.69     & .41&0.  &.0609&.275  &.122  &3.97&          &            &      \\
squalane         &170  &1.26& 3.08 &.5   &.77     & .43&0.  &.0755&.275  &.122  &4.39&0.015     &  0.2       &.23   \\
squalane         &172  &1.25& 1.00 &.5   &.82     & .41&0.  &.0916&.275  &.122  &4.70&0.0148    & -0.24      &.30   \\
squalane         &174  &1.22&-0.82 &.5   &.86     & .41&0.  &.1047&.275  &.122  &4.97&          &            &      \\
squalane         &176  &1.18&-2.37 &.5   &.89     & .41&0.  &.1117&.275  &.122  &5.12&          &            &      \\
\hline
hydrogen bonding       &&&&&&&&&&&&&& \\
glycerol         &192 &4.12 & 3.55 &.53  &1.59    &.50 &0.68&.0063&.336  &.211  &13.4&          &            &      \\
glycerol         &194 &4.03 & 2.41 &.53  &1.55    &.49 &0.68&.0071&.336  &.211  &13.1&          &            &      \\
glycerol         &196 &3.95 & 1.35 &.54  &1.53    &.48 &0.68&.0082&.336  &.211  &13.0&          &            &      \\
glycerol         &198 &3.87 & 0.37 &.54  &1.54    &.48 &0.68&.0092&.336  &.211  &13.0&          &            &      \\
glycerol         &200 &3.79 &-0.54 &.55  &1.58    &.49 &0.68&.0102&.336  &.211  &13.3&          &            &      \\
propylene glycol &171 &4.05 & 1.26 &.54  &1.10    &.37 &0.64&.0118&.32   &.176  &8.2 &          &            &      \\
propylene glycol &174 &3.83 &-0.36 &.58  &1.22    &.40 &0.64&.0172&.32   &.176  &8.9 &          &            &      \\
propylene glycol &177 &3.67 &-1.85 &.60  &1.31    &.42 &0.64&.0231&.32   &.176  &9.5 &          &            &      \\
propylene glycol &180 &3.54 &-3.20 &.64  &1.40    &.44 &0.64&.0323&.32   &.176  &10.2& 62.9     &-3.04       & 0.5  \\
\hline
		\end{tabular}
	\caption{Model parameters fitted to shear relaxation data of glass formers with pronounced secondary relaxation peak, including two hydrogen bonding substances: $G$ shear modulus, $\tau_c$ terminal relaxation time, $\beta$ Kohlrausch exponent, $f_r$ prefactor for the reversible relaxations, $f_{r0}$ recalculated $f_r$, supposed to be close to the theoretical value 0.4409, $p_r$ return probability in the irreversible region, amp, pos and fwhm describe the gaussian distribution of secondary relaxations. $GJ_0$ is the resulting recoverable compliance. $\Delta\epsilon$ is the strength, $v_\epsilon$ the cutoff parameter, and $f_\epsilon$ the additional secondary relaxation factor for the dielectric relaxation. References for the data: dibutyl phtalate shear \cite{maggi} and dielectric \cite{albena}; squalane \cite{niss}; glycerol \cite{glynew}; propylene glycol shear \cite{maggi} and dielectric \cite{albena}.}
	\label{tab:rse}
\end{table*}

\begin{figure}   
\hspace{-0cm} \vspace{0cm} \epsfig{file=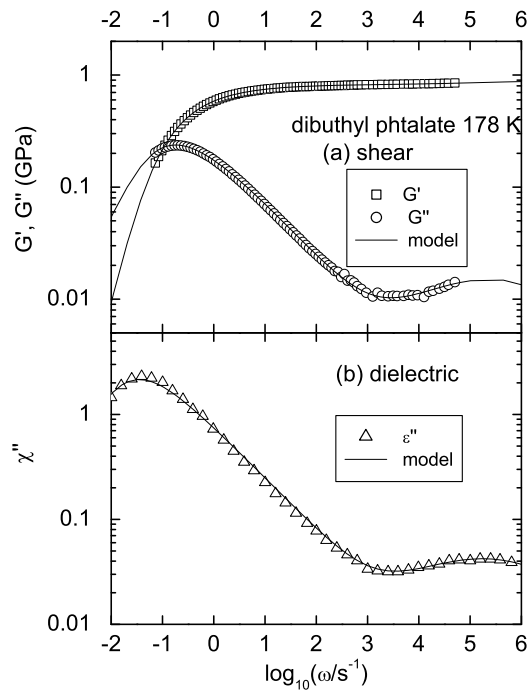,width=7 cm,angle=0} \vspace{0cm} \caption{(a) Dynamic shear data \cite{maggi} (b) dielectric data \cite{albena} of dibutyl phtalate at 178 K fitted with the parameters in Table II.}
\end{figure}

In glass formers with a strong secondary relaxation peak, one always finds a larger $f_r$ than the theoretical value 0.4409.

This is illustrated in Table II for two cases, dibutyl phtalate and squalane.

The secondary relaxation peak in dibutyl phtalate is not very strong, but clearly visible (see Fig. 5 (a)); the one in squalane is fairly strong (see Fig. 6 (a)).

Figs. 5 (b) and 6 (b) show the corresponding fits of dielectric data at the same temperature for the same substance. It turns out that the strength $f_\epsilon amp$ of the secondary peak is about the same as for the shear in dibutyl phtalate, but a factor three to four weaker than the shear one in squalane.  

\begin{figure}   
\hspace{-0cm} \vspace{0cm} \epsfig{file=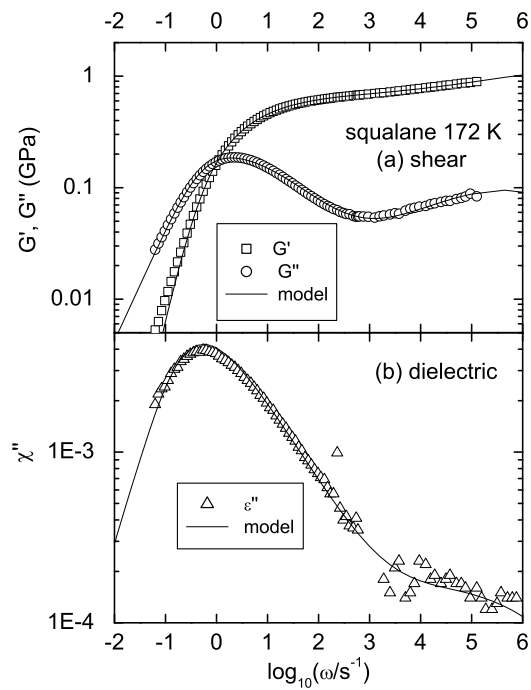,width=7 cm,angle=0} \vspace{0cm} \caption{(a) Dynamic shear data \cite{niss} (b) dielectric data \cite{niss} of squalane at 172 K fitted with the parameters in Table II.}
\end{figure}

\begin{figure}  
\hspace{-0cm} \vspace{0cm} \epsfig{file=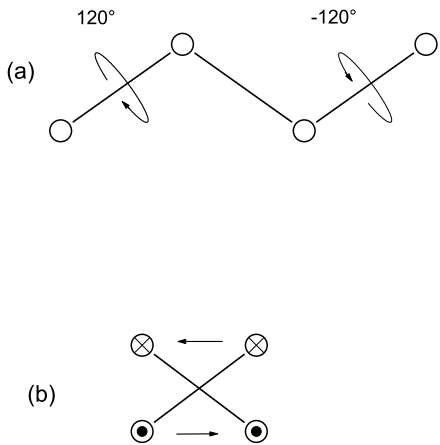,width=6 cm,angle=0} \vspace{0cm} \caption{(a) The Helfand crankshaft motion of a {\it trans}-conformation of four carbon atoms consists of a 120 degree rotation of the first carbon bond and a -120 degree counter rotation of the last carbon bond, leading again to a {\it trans} conformation (b) The shear motion of the two chain ends in the plane perpendicular to the rotating bonds.}
\end{figure}

Let us describe the peak by adding a gaussian barrier distribution $l_G(v)$ to $l_r(v)$ in eq. (\ref{lv}), centered at the position pos (in eV), with a full width at half maximum fwhm, also in eV, and with the amplitude amp. The $\tau$-values belonging to a given barrier are calculated with eq. (\ref{arrh}).

With increasing temperature, the amplitude increases strongly in both substances (see Table II), similar to the behavior of the secondary relaxation peak in tripropylene glycol observed earlier \cite{olsen}.

The astonishing finding is that $f_r$ does also increase, with the increase of the secondary peak. $f_r$ follows the equation
\begin{equation}\label{fr0}
	f_r=f_{r0}+\int_{-\infty}^\infty l_G(v)dv,
\end{equation}
where $f_{r0}$ stays constant and equal to the theoretical value 0.4409 within experimental error.

This finding suggests the following interpretation: The slower reversible Eshelby transitions absorb the faster secondary processes. Their eigenvector then is no longer a pure Eshelby process, but begins to contain secondary eigenvector contributions.

If the secondary relaxation changes the shape or the orientation of the molecule, it couples to an external shear stress. But it does not contribute to the viscous flow. Sooner or later, the process is reversed and the gain in shear strain is lost.

This is principally different for the Eshelby processes, which can break and reform nearest-neighbor bonds and which do contribute to the viscous flow.

But if they begin to contain an eigenvector component from molecular shape or orientation changes, their contribution to the viscous flow in the irreversible region diminishes accordingly.

This is most clearly seen in squalane. Squalane is a short polymer of 24 carbon atoms, with occasional CH$_3$ side groups which help to avoid crystallization.

In such a polymer, the viscous flow induces torsional rotations around C-C bonds. However, a carbon-carbon bond in the middle of the chain cannot swing two long chain ends around, so the motion of a polymer in the melt requires correlated rotational jumps around at least two carbon-carbon bonds, preferably parallel ones to keep the orientation of the chain ends constant.

When the problem became accessible to numerical treatment, Helfand \cite{helfand} found that in most cases there is just one carbon-carbon bond between the two parallel counter rotating ones, limiting the resulting Helfand crankshaft to four carbon atoms (see Figure 7 (a)).

In the Helfand case, the two rotating and counter rotating bonds are parallel, but they are not collinear. This implies a displacement of the two chain ends with respect to each other, a displacement which is perpendicular to the direction of the two rotating bonds (see Figure 7 (b)).

In this Helfand picture, one understands why the dielectric reaction is so much weaker, because one shears a unit of four carbon chain atoms, but only rotates one C-C bond. One also understands the average barrier height of 0.275 eV, because it is about twice the torsional rotation barrier \cite{tsuzuki,allinger}.

It is clear that the absorption of these Helfand processes into the Eshelby eigenvectors leads to a lower contribution to the viscous flow.

It should be mentioned that an electric-circuit-equivalent fit has been made to new squalane shear data \cite{squanew}. It is in several ways parallel to the present treatment, but the relation between the electric-circuit parameters and the ones in Table II is not straightforward. 

Looking back to the (SiO$_2$)$_{75}$(Na$_2$O)$_{25}$ data in Fig. 4, the question arises: Why is there no enhancement of $f_r$ in this substance, which has two secondary relaxation peaks and a lot of fast relaxations in the bargain?

The answer is that $l_r(v)dv$ is defined in terms of the recoverable compliance $\delta J$ in units of $1/G$, where $G$ is the shear modulus extrapolated from the relaxations in the measurement window. If there are faster relaxations outside the window, they remain invisible in $f_r$, because one does not know that the unit $1/G$ is smaller than one supposes.

In fact, the discussion will show that this is always the case, and that the $G$-values of Table I and II are not really an infinite frequency shear modulus, but rather a microsecond shear modulus.

\subsection{Hydrogen bonding glass formers}

To describe the glycerol shear relaxation in Fig. 1, one has to modify the model in one crucial point: One has to assign to the irreversible Eshelby processes not the return probability $p_r=0$, but rather a nonzero value.

To keep the sample flowing, one then needs an increase of the Eshelby density by the factor $1/(1-p_r)$.

In the two examples in Table II, glycerol \cite{glynew} and propylene glycol \cite{maggi}, one also has to take a secondary relaxation peak into account.

Fortunately, the position and width of the gaussian distribution of these secondary relaxation peaks are known. For glycerol, one can take them from dielectric glass phase data \cite{gainaru}. For propylene glycol, one can take them from dielectric tripropylene glycol data \cite{olsen}, because this peak develops from an excess wing in propylene glycol to a pronounced secondary relaxation peak in the polypropylene glycols \cite{ppgbeta}.

With the introduction of a nonzero return probability in the irreversible range, eq. (\ref{fr0}) changes to
\begin{equation}\label{fr0pr}
	f_r=f_{r0}/(1-p_r)+\int_{-\infty}^\infty l_G(v)dv,
\end{equation}
where $f_{r0}$ should be again the theoretical value 0.4409 to enable the viscous flow.

\begin{figure}   
\hspace{-0cm} \vspace{0cm} \epsfig{file=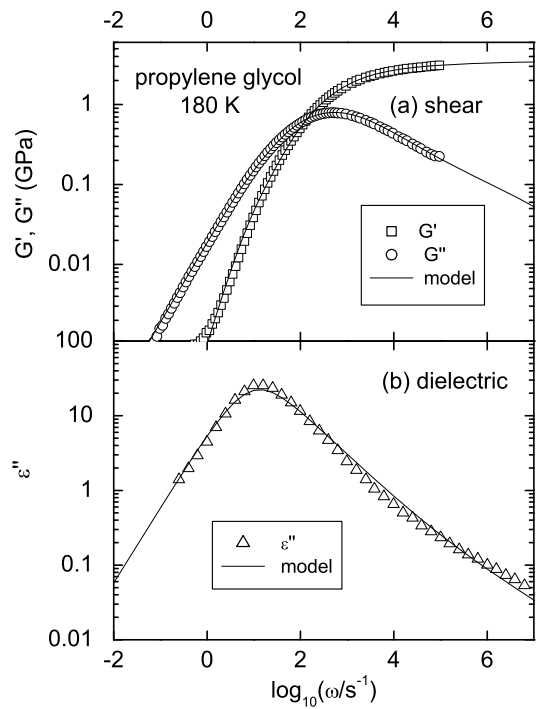,width=7 cm,angle=0} \vspace{0cm} \caption{(a) Dynamic shear data \cite{maggi} (b) dielectric data \cite{albena} of propylene glycol at 180 K fitted with the parameters in Table II.}
\end{figure}

Fig. 8 shows a fit of shear \cite{maggi} and dielectric \cite{albena} data in propylene glycol at 180 K in terms of the parameters in Table II.

The fit of the glycerol shear data \cite{glynew} at 196 K is shown in Fig. 1. For both substances, one finds $f_{r0}$-values close to the theoretical one.

But if one tries the fit of Fig. 8 (b) for the many dielectric glycerol data in the literature, it does not work; the theoretical peak is always much broader than the measured ones.

To see whether the irreversible Eshelby prediction for the dynamic heat capacity holds, the $\tau_c$ values of the shear data in Table II were again fitted with a Vogel-Fulcher law. With this $\tau_c$ temperature dependence and the $\Delta c_p$-values given in the caption of Fig. 9, one can calculate the temperature dependence of $c_p''$ for a given frequency.

\begin{figure}   
\hspace{-0cm} \vspace{0cm} \epsfig{file=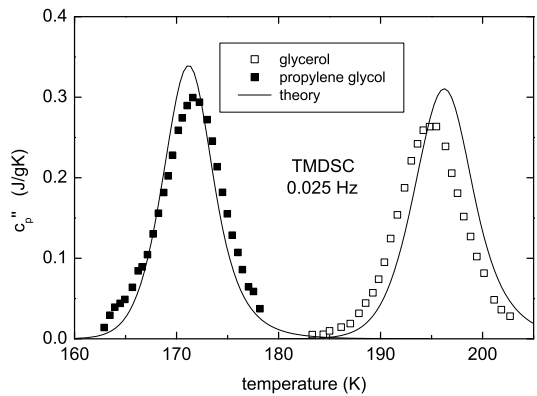,width=7 cm,angle=0} \vspace{0cm} \caption{Temperature modulated differential scanning data \cite{carpentier} for the imaginary part $c_p''$ of the dynamic heat capacity of glycerol and propylene glycol. The continuous line is calculated from eq. (\ref{pt}) for the irreversible Eshelby processes, using $\Delta c_p=0.75$ J/gK for glycerol and $\Delta c_p=0.82$ J/gK for propylene glycol. The $\tau_c$ temperature dependence was taken from the data in Table II.}
\end{figure}


The result is compared to experimental data in glycerol and propylene glycol \cite{carpentier} in Fig. 9 and supports again the validity of the theoretical consideration \cite{asyth}, this time in two hydrogen bonding substances.

\section{Discussion and conclusions}

Let us begin the discussion with the microscopic reason for the fragility, attributed by Adam-Gibbs \cite{adam} to the increasing size of the cooperatively rearranging regions and by the elastic models \cite{nemilov,dyre,nelson,shov2015} to the temperature dependence of the short time shear modulus $G$. Both concepts have experimental support \cite{bauer,nelson}.

The first thing to note in this context is that one should wonder about the strong temperature dependence of $G$.

Let us take the well-studied example of selenium, where $G\propto 1/T^4$ according to ultrasonic data in the undercooled liquid \cite{reichardt}. The exponent 4 is close to the value 4.4 required for the validity of the elastic model \cite{bzr}.

Using the definition of the Gr\"uneisen parameter $\gamma=-\partial\ln{\omega}/\partial\ln{V}$  and the thermal volume expansion coefficient $\alpha=3.63\ 10^{-4}$ per degree Kelvin \cite{simha}, one arrives at a shear wave Gr\"uneisen parameter of 18.1, much higher than the average Gr\"uneisen parameter between 1 and 2 measured in the glass phase \cite{meissner}.

Obviously, the strong decrease of $G$ is not due to the thermal expansion. What then is its reason?

A part of the reason has been clarified in neutron scattering measurements \cite{bzr,hansen}: the boson peak softens strongly with increasing temperature in the undercooled liquid, leading to a strong enhancement of the mean square displacement.

The strong influence of the boson peak vibrations on the shear modulus has been demonstrated convincingly in numerical simulations \cite{leonforte} of a Lennard-Jones glass, where the real (nonaffine) shear modulus was a factor of two smaller than the hypothetical affine one.

But it is not only the boson peak. Taking only the mean square displacement of the motions in the picosecond range, one needs to subtract the crystalline mean square displacement before one achieves a successful description of the viscosity \cite{bzr}.

A better agreement is obtained by measuring the mean square displacement with a backscattering instrument on the nanosecond time scale \cite{hansen}. In this case, one needs no subtraction to obtain agreement with the measured shear modulus and the measured fragility.

This provides the answer to the question: the short time shear modulus $G$ fitted to the shear data is a microsecond shear modulus, influenced not only by the boson peak vibrations, but also by all reversible relaxations with relaxation times shorter than a microsecond. That is the reason for its strong temperature dependence.

This insight is supported by very recent simulations \cite{lerner}, which found a strong temperature dependence for the restoring forces against a dipole force pair on neighboring atoms, but only a weak temperature dependence for the zero Kelvin shear modulus. 

Realizing this, one realizes that the proportionality of $G$ to the flow barrier \cite{nemilov,dyre,nelson,shov2015} must have a deeper reason than a simple scaling of the energy landscape with $G$, and also that this proportionality does not necessarily rule out the Adam-Gibbs explanation \cite{adam}.

In fact, it will be seen below that the results of the present paper suggest the validity of the Adam-Gibbs postulate. 

The existence of a sizable relaxation contribution between a microsecond and a picosecond has been firmly established by many experiments in many glass formers, both in the glass \cite{gilroy} and the undercooled liquid. It grows with increasing temperature even in the glass phase (this is evidenced in Fig. 3 by the Brillouin data at 25 and 35 GHz, see the detailed discussion in reference \cite{philmag2002}).

From our finding that the Eshelby density in the internal friction is the same as the one fitted at the glass temperature $T_g$, it follows that it is not temperature-dependent in the glass phase. The changes occur at smaller barriers.

The growth at small barriers becomes dramatic in the undercooled liquid. This is generally seen from the strong increase of the light scattering Brillouin damping with increasing temperature above $T_g$, and has been shown quantitatively for polystyrene from Brillouin and neutron scattering data \cite{burel}.

From all this, and from the demonstration in Section III. C that the slow relaxations pick up some of the eigenvectors of the faster ones, it is very clear that the supposedly pure shear distortion of the Eshelby transitions of reference \cite{asyth} is not really a pure shear, but interspersed with nonaffine atomic motions and low barrier relaxations.

On the other hand, the boson peak vibrations and the fast relaxations have already done their reduction of the shear modulus when one reaches the millisecond range. The elastic shear energy of a state with a given elastic strain and the lifetime of the terminal relaxation time has to be calculated with $G$. For this reason, the calculation of the relaxation time distribution of the irreversible Eshelby states \cite{asyth}, with the microsecond shear modulus $G$ as a basis, is consistent.

The assumption of the present paper, namely a Kohlrausch density of Eshelby states going with a constant Kohlrausch exponent $\beta$ through the terminal relaxation time, can be checked by the $f_r$ values which one finds. The truth of the assumption is supported by the three metallic glasses and the four van der Waals molecular glass formers in Table I.

The network glass formers do not all support the assumption, but this  might be due to bond breaking processes shortly above $\tau_c$. In silica, one can check the fit results against torsion pendulum or other internal friction data in the glass phase, and finds agreement within experimental error.

The van der Waals molecular glass formers with a secondary relaxation do not only support the assumption, but show clearly that the Eshelby relaxations incorporate eigenvector contributions from the secondary relaxation states. 
In the hydrogen bonding substances, one has to postulate a strong nonzero return probability in the irreversible relaxation time region. Then one gets again consistence with the continuity assumption for the Kohlrausch density.
It is tempting to connect this nonzero return probability to the long lifetime of hydrogen-bonded structures in the mono alcohols \cite{mono}, but this goes beyond the scope of the present paper.

Having established the validity of the Kohlrausch Ansatz in many examples, let us proceed to its theoretical interpretation.

For a transforming Eshelby domain consisting of $N$ particles, the Kohlrausch $\beta$ results from the increase of the number of structural possibilities with increasing $N$, combined with the increasing barrier height with increasing $N$. Let $S_1$ be the structural entropy per particle, $V_1$ the increase of the barrier $V_c$ per particle. Then
\begin{equation}
	\beta=\frac{S_1T}{V_1}.
\end{equation}

$S_1$ decreases to zero at the Kauzmann temperature, so one expects a proportionality of $S_1T$ to about $T^4$ for the average van der Waals fragility. But in the examples of DC704 and PPE in Table I, $\beta$ does change only a few percent for a temperature change of five percent (in PPE even in the wrong direction), where it should decrease by twenty percent for a temperature-independent $V_1$.

Obviously, $V_1$, which one expects to increase by the anharmonicity, does instead decrease with decreasing temperature. How is this possible?

A plausible answer to this question is that the Eshelby transformation starts by a shear banding transition, a planar transition in the middle of the Eshelby sphere, with a barrier $V_c$ which is not proportional to $N$, but rather to $N^{2/3}$. With this assumption,
\begin{equation}
	V_1=\frac{\partial V_c}{\partial N}=\frac{2V_c}{3N}
\end{equation}
and
\begin{equation}\label{beta}
	\beta=\frac{3NS_1T}{2V_c}.
\end{equation}
In this equation, the decrease of $S_1T$ is compensated by the increase of $N$.

Eq. (\ref{beta}) allows to calculate $N$ from measurable quantities. 

To take an example: the excess entropy per atom of the metallic glass vitralloy-1 over the crystal \cite{busch} close to $T_g$ is 0.36 $k_B$. Since the boson peak of vitralloy is not very anharmonic \cite{schober}, one can neglect the vibrational part of the excess entropy and identify this value with $S_1$. With $V_c\approx 35k_BT_g$ and $\beta=0.431$ from Table I, this leads to $N=27.9$ atoms in the crossover Eshelby domain in vitralloy at its glass transition. 

The last point to discuss concerns the thermal properties of undercooled liquids.

A convincing proof of the scheme of this and the previous paper \cite{asyth} is the quantitative predictability of dynamic heat capacity results on the basis of dynamic shear data. This works not only for a metallic glass (see Figure 2) and a van der Waals glass former \cite{asyth}, but also for the two hydrogen bonding examples in Fig. 9.

The finding implies that reversible transitions are not seen in the dynamic heat capacity.

Naturally, the contribution of a reversible transition to the structural entropy is not zero. But this entropy relaxes only when the reversible transition is created or destroyed. This happens when a larger domain in the immediate neighborhood makes an irreversible jump.

To conclude, the paper extends exact results for the highly viscous flow by irreversible Eshelby translations to reversible Eshelby translations by assuming a continuous Kohlrausch density of the Eshelby transitions at the terminal relaxation time. The model describes successfully a wide variety of experimental data. 

The weak temperature dependence of the fitted Kohlrausch parameters cannot be understood in terms of a proportionality of the terminal barrier to the number of particles in the Eshelby domain; one rather has to postulate a shear banding planar transition in the middle of the domain (at least as the initial step determining the barrier height).

With this postulate, the model encourages the hope for a unified picture of the many puzzling properties of glass forming liquids close to their glass transition.

\end{document}